\documentclass[10pt]{article}
\usepackage[dvips]{graphicx}
\usepackage{amsmath}
\newcommand{\bi}[1]{\textbf{\textit{#1}}}

\begin{document}
\title{\textbf{Charge screening in single-walled armchair carbon nanotubes}}
\author{Vadym M. Adamyan\footnote{E-mail: vadamyan@paco.net}, Oleksii A. Smyrnov\footnote{E-mail: smyrnov@onu.edu.ua}\\
\emph{Department of Theoretical Physics, Odessa I. I. Mechnikov National University,}\\
\emph{2 Dvoryanskaya St., Odessa 65026, Ukraine}}
\date{February 5, 2007}
\maketitle
\begin{abstract}
The screened effective quasione-dimensional interaction potential
for charged particles localized on the wall of a single-walled
metallic armchair carbon nanotube (CNT) was obtained within the
framework of the random phase approximation (RPA).
\end{abstract}

PACS number(s): 73.63.Fg

\section{Introduction}
\setcounter{equation}{0}

The single-walled CNT is a strip of the 2D graphite plane rolled
up into cylinder. Remind that the 2D graphite plane is composed of
hexagons, every vertex of which is the carbon atom. The chirality
of CNT is defined by two integer numbers $(n, m)$ which specify
the 2D co-ordinate of the hexagon imposed on the hexagon located
at the origin, when the strip of the 2D graphite layer is rolled
up. CNTs with $(n, n)$ chirality are the "armchair" CNTs. All the
previous works \cite{ando}-\cite{dresselh2} clearly assert that
armchair CNTs with any chirality numbers have metallic
conductivity, that is why here we can use approximations that
proper to the theories of metals. However, fundamental results of
the theory of metals cannot be applied directly to armchair
nanotubes for their one-dimensional structure. At the same time
their one-dimensional analogues are of the barest necessity for
investigation of optical and transport properties of nanotubes.
This note is aimed at derivation of one important result of this
kind, namely the derivation of explicit expression for the
effective quasione-dimensional potential of charged interstitial
defect, which would account the collective effect of dielectric
screening of the defect by free electrons of nanotube. Because of
its relatively large length $(\sim 1-10~\mathrm{\mu m})$ and small
diameter $(\sim 1-10~\mathrm{nm})$ CNT may be considered as a
quasione-dimensional system. So we've used here the Lindhard
screening theory (so-called RPA) in one dimension. Then in the
limiting case of small wavenumber we have got the Thomas-Fermi
screened potential for two charges localized on the wall of tube.
The further transverse integration of the obtained expression
yields the self-consistent one-dimensional charge-charge
interaction potential, which can be used for account of
many-particle effects in CNTs.

\section{General form of screened potential}
\setcounter{equation}{0}

Within the framework of the Lindhard screening theory we consider
the one-dimensional Fourier transform of the following Poisson
equation for screened electrostatic potential $\varphi(\bi{r})$:
\begin{equation}\label{2.1}
\left(q^2-\Delta_\mathrm{2D}\right)\varphi(q,\bi{r}_\mathrm{2D})=4\pi\left(\rho^\mathrm{ext}(q,\bi{r}_\mathrm{2D})+\rho^\mathrm{ind}(q,\bi{r}_\mathrm{2D})\right),
\end{equation}
where $\rho^\mathrm{ext}(q,\bi{r}_\mathrm{2D})$ is the
one-dimensional Fourier transform along the tube axis of the
density of extraneous charge,
$\rho^\mathrm{ind}(q,\bi{r}_\mathrm{2D})$ is that of the charge
density induced by the extraneous charge, $q$ is the longitudinal
component of wave vector, and $\bi{r}_\mathrm{2D}$ is the
transverse component of the radius-vector. For simplicity we will
assume that $\rho^\mathrm{ext}$ is axial symmetric,
$\rho^\mathrm{ext}(q,\bi{r}_\mathrm{2D})=\rho^\mathrm{ext}(q,r_\mathrm{2D})$,
and localized in the small vicinity of the tube wall. As follows
$\varphi(q,\bi{r}_\mathrm{2D})$,
$\rho^\mathrm{ind}(q,\bi{r}_\mathrm{2D})$ depend on
$\bi{r}_\mathrm{2D}$ only through $r_\mathrm{2D}$ and besides
whatever the case $\rho^\mathrm{ind}$ is localized in the small
vicinity of the tube walls. The $\rho^\mathrm{ind}$ may be written
as follows:
\begin{equation}\label{2.2}
\rho^\mathrm{ind}(q,r_\mathrm{2D})=-e\int\limits^L_0\exp(-\mathrm{i}qz)\sum_{k,s}f(E_s(k))|\Psi_{k,s}(\bi{r})|^2\mathrm{d}z-\rho^0(q,r_\mathrm{2D}),
\end{equation}
where $f$ is the Fermi-Dirac function, $E_s(k)$ and
$\Psi_{k,s}(\bi{r})=\exp(\mathrm{i}kz)u_{k,s}(\bi{r})/\sqrt{N}$
are the band energies and the corresponding Bloch wave functions
of $\pi$-electrons, $\rho^0$ is the charge density without
extraneous charges and $L$ is the length of CNT. Following the
Lindhard method we get the linear in $\varphi$ approximation for
the induced charge density:
\begin{equation}\label{2.3}
\begin{split}
&\rho^\mathrm{ind}(q,r_\mathrm{2D})=-e^2\sum_{k,s}\sum_{k',s'}\frac{f(E_{s'}(k'))-f(E_s(k))}{E_s(k)-E_{s'}(k')}\\
&\times\int\limits^L_0\frac{1}{N^2}\int\limits_{\mathrm{E_3}}u^*_{k,s}(\bi{r})u_{k',s'}(\bi{r})\varphi(\bi{r})\exp(\mathrm{i}z(k'-k))\mathrm{d}\bi{r}u^*_{k',s'}(\bi{r})u_{k,s}(\bi{r})\exp(\mathrm{i}z(k-k'-q))\mathrm{d}z,
\end{split}
\end{equation}
where $N$ is the number of unit cells in CNT and $s,s'$ number
$\pi$-electrons bands crossed by the Fermi level. Further, writing
$\varphi(\bi{r})$ in the form
$$
\varphi(\bi{r})=\frac{1}{L}\sum_q\exp(\mathrm{i}qz)\varphi(q,r_\mathrm{2D})
$$
and taking into account that the sought potential varies slightly
inside the unit cell we get from~(\ref{2.3}) that
\begin{equation}\label{2.4}
\begin{split}
\rho^\mathrm{ind}(q,r_\mathrm{2D})&=-\frac{e^2}{L}\sum_{k,s,s'}\frac{f(E_s(k))-f(E_{s'}(k-q))}{E_{s'}(k-q)-E_s(k)}\\
&\times\varphi(q,R)B_{ss'}(k,k-q,a)\int\limits^a_0u^*_{k-q,s}(z,
\bi{r}_\mathrm{2D})u_{k,s'}(z,\bi{r}_\mathrm{2D})\mathrm{d}z,
\end{split}
\end{equation}
where $a$ is the longitudinal period of nanotube, $R$ - its
radius, and
\begin{equation}\label{2.5}
B_{ss'}(k,k-q,a)=\int\limits_\mathrm{E_2}\int\limits^a_0u^*_{k,s'}(z,
\bi{r}_\mathrm{2D})u_{k-q,s}(z,\bi{r}_\mathrm{2D})\mathrm{d}z\mathrm{d}\bi{r}_\mathrm{2D}.
\end{equation}
By~(\ref{2.1})
\begin{equation}\label{2.6}
\varphi(q,r_\mathrm{2D})=4\pi\int\limits_\mathrm{E_2}G_0(q,\bi{r}_\mathrm{2D},\bi{r}'_\mathrm{2D})\left(\rho^\mathrm{ext}(q,r'_\mathrm{2D})+\rho^\mathrm{ind}(q,r'_\mathrm{2D})\right)\mathrm{d}\bi{r}'_\mathrm{2D},
\end{equation}
where $G_0(q,\bi{r}_\mathrm{2D},\bi{r}'_\mathrm{2D})$ is the Green
function for the 2D Helmholtz equation,
\begin{equation}\label{2.7}
G_0(q,\bi{r}_\mathrm{2D},\bi{r}'_\mathrm{2D})=\frac{1}{2\pi}K_0(|q||\bi{r}_\mathrm{2D}-\bi{r}'_\mathrm{2D}|),
\end{equation}
where $K_0(|q||\bi{r}_\mathrm{2D}-\bi{r}'_\mathrm{2D}|)$ is the
modified Bessel function of the second kind. By our assumptions
both the $\rho^\mathrm{ext}(q, r_\mathrm{2D})$ and
$\rho^\mathrm{ind}(q, r_\mathrm{2D})$ are actually localized on
the tube wall. Hence, for $r_\mathrm{2D}=R$ from~(\ref{2.6})
and~(\ref{2.7}) we have
\begin{equation}\label{2.8}
\varphi(q, R)=\varphi^\mathrm{ext}(q,
R)+2K_0(|q|R)I_0(|q|R)\int\limits_\mathrm{E_2}\rho^\mathrm{ind}(q,
r'_\mathrm{2D})\mathrm{d}\bi{r}'_\mathrm{2D},
\end{equation}
where $\varphi^\mathrm{ext}(q, R)$ is the contribution to the
Fourier transform of the total potential from $\rho^\mathrm{ext}$,
and $I_0(|q|R)$ is the modified Bessel function of the first kind.
We see from~(\ref{2.4}) and~(\ref{2.8}) that
\begin{equation}\label{2.9}
\begin{split}
&\varphi(q, R)=\frac{\varphi^\mathrm{ext}(q,R)}{\varepsilon_R(q)},\\
&\varepsilon_R(q)=1+\frac{e^2}{\pi}\sum_{s,s'}\int\limits^{\pi/a}_{-\pi/a}\frac{f(E_s(k))-f(E_{s'}(k-q))}{E_{s'}(k-q)-E_s(k)}|B_{ss'}(k,k-q,a)|^2\mathrm{d}k
I_0(|q|R)K_0(|q|R).
\end{split}
\end{equation}
For low and room temperatures the main contributions
to~(\ref{2.9}) are made by quasi-momenta from the small vicinity
of the Fermi quasi-momentum $k_\mathrm{F}$. Due to the
orthogonality and normalization condition of the Bloch functions
for $k\simeq k-q\simeq k_\mathrm{F}$ we have
\begin{equation}\label{2.10}
|B_{ss'}(k,k-q,a)|^2\simeq\left\{\begin{array}{c}0~\mathrm{for}~s\neq s',\\
1~\mathrm{for}~s=s',\\
\end{array}\right.
\end{equation}
and, besides with account of the spin degeneracy
\begin{equation}\label{2.11}
\frac{f(E_s(k))-f(E_{s'}(k-q))}{E_{s'}(k-q)-E_s(k)}\simeq2\delta(E_{s'}(k)-E_\mathrm{F}),
\end{equation}
where $\delta(E_{s'}(k)-E_\mathrm{F})$ is the Dirac
delta-function. It follows from~(\ref{2.9})-(\ref{2.11}) that
\begin{equation}\label{2.12}
\varepsilon_R(q)=1+\frac{2e^2}{\pi\hbar}\sum_s\frac{1}{V_s}I_0(|q|R)K_0(|q|R),
\end{equation}
where
\begin{equation}\label{2.13}
V_s=\frac{1}{\hbar}\left|\frac{\partial E_s(k)}{\partial
k}\right|_{k=k_\mathrm{F}}
\end{equation}
is the velocity of electrons of the $s$-th band on the Fermi
level.

Actually, according to~(\ref{2.9}) $\varepsilon_R(q)$ is an
analogue of the Thomas-Fermi dielectric function for any
quasione-dimensional metallic nanotube. The screened
quasione-dimensional electrostatic potential produced by a charge
$e_0$, distributed with the density
$$
\rho^\mathrm{ext}(\bi{r})=\frac{e_0}{2\pi
R}\delta(r_\mathrm{2D}-R)\delta(z),
$$
in accordance with~(\ref{2.12}) is given by:
\begin{equation}\label{2.14}
\varphi(z,R)=\frac{e_0}{\pi
R}\int\limits^{\infty}_{-\infty}\frac{I_0(|q|)K_0(|q|)\exp(\mathrm{i}
qz/R)}{1+gI_0(|q|)K_0(|q|)}\mathrm{d}q
\end{equation}
with the constant
\begin{equation}\label{2.15}
g=\frac{2e^2}{\pi\hbar}\sum_s\frac{1}{V_s}.
\end{equation}
Note that the interaction energy $E_\mathrm{q}$ of electrons of
infinite nanotube with the given external charge due to screening
appears to be finite:
\begin{equation}\label{2.16}
E_\mathrm{q}=\frac{2ee_0}{g}n_\mathrm{e},
\end{equation}
where $n_\mathrm{e}$ is the number of electrons per unit length of
nanotube.

\section{Numerical calculations}
\setcounter{equation}{0}

Both the velocity of electrons and Fermi energy for the $(n, n)$
carbon nanotube can be obtained from any single-electron model
(the tight binding method, for example). The band structure of 2D
graphite was obtained in~\cite{wallace} within the framework of
the tight binding method. It has up to constant shift the
following form:
\begin{equation}\label{3.1}
E(k_x,k_z)=\pm\gamma_0\sqrt{1+4\cos\left(\frac{\sqrt{3}k_xb}{2}\right)\cos\left(\frac{k_zb}{2}\right)+4\cos^2\left(\frac{k_zb}{2}\right)},
\end{equation}
where $\gamma_0=2.79~\mathrm{eV}$ and $b=0.246~\mathrm{nm}$ are
the nearest-neighbour transfer integral and the in-plain lattice
constant, respectively. According to~\cite{dresselh2}, when a 2D
graphite layer is rolled up as a cylinder the number of allowed
states in the circumferential direction becomes limited. So
allowed values for the wavenumber in the circumferential direction
for armchair nanotubes can be written as:
\begin{equation}\label{3.2}
k^\nu_x=\frac{\nu}{n}\frac{2\pi}{b\sqrt{3}},
\end{equation}
where $\nu=1~\ldots~n$, $n$ is the chirality number.
\begin{figure}[t]
\begin{center}
\includegraphics[scale=0.7]{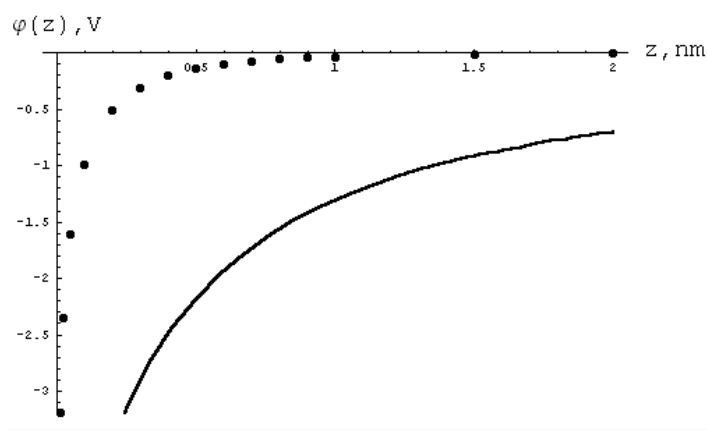}
\caption{\label{fig1}The screened potential $\varphi(z)$ according
to~(\ref{2.14}) for the (5, 5) nanotube (plotted point by point)
in comparison with the unscreened averaged Coulomb potential
(solid line)}
\end{center}
\end{figure}
Thus the one-dimensional dispersion relation for $2n$ energy bands
of the $(n, n)$ single-walled armchair CNT was obtained
in~\cite{dresselh2}. It follows from~(\ref{3.1}) and~(\ref{3.2})
that with respect to~(\ref{3.1}) the Fermi energy $E_\mathrm{F}=0$
and only two bands
$$
E_\pm(k)=\pm\gamma_0\left(1-2\cos\left(\frac{kb}{2}\right)\right)
$$
cross the Fermi level at $k_\pm=k_\mathrm{F}=2\pi/3b$. As follows,
$$
V_\pm=\frac{\gamma_0b\sqrt{3}}{2\hbar}
$$
and
\begin{equation}\label{3.3}
g=\frac{8e^2}{\pi\sqrt{3}\gamma_0b}
\end{equation}
is the universal constant for all the armchair tubes within the
framework of the tight binding approximation. Figure 1 shows the
numerically calculated screened potential~(\ref{2.14}) for the (5,
5) nanotube, with radius $R=0.339~\mathrm{nm}$, in comparison with
the unscreened averaged over axial component Coulomb potential.

Thus the dielectric screening of the Coulomb potential by free
$\pi$-electrons in armchair nanotubes results in substantial
reducing of the effective depth of Coulomb well at the origin and
rather faster vanishing of the corresponding potential at
infinity.

\end{document}